\documentclass{aa}

\usepackage{graphics}
\usepackage{amssymb}
 
\begin{document}

\thesaurus{06     
              (08.06.2  
               09.03.1;  
               09.09.1;  
               09.10.1;  
               09.13.2)} 

\title{L483: a protostar in transition from Class 0 to Class I}

\author{
	M. Tafalla\inst{1}
\and
        P.C. Myers\inst{2}
\and
        D. Mardones\inst{3}
\and
        R. Bachiller\inst{1}
         }

\offprints{M. Tafalla}

\institute{
        Observatorio Astron\'omico Nacional, Apartado 1143, E-28800 Alcal\'a de 
        Henares, Spain\\
        email: tafalla@oan.es, bachiller@oan.es
\and
        Harvard-Smithsonian Center for Astrophysics, MS 42, 60 Garden St.,
        Cambridge, MA 02138, USA\\
        email: pmyers@cfa.harvard.edu
\and
        Departamento de Astronom\'{\i}a, Universidad de Chile,
        Casilla 36-D, Santiago, Chile\\
        email: mardones@das.uchile.cl
            }

\date{Received 14 Oct. 1999; accepted 24 May 2000}

\maketitle

\begin{abstract}

We present line observations of different molecular species toward 
the dense core in \object{L483} and
its bipolar outflow powered by the Class 0 object \object{IRAS 18148-0440}. 
$^{12}$CO(2--1) maps show that the outflow is well collimated and asymmetric, 
and that its gas is warmer than the surrounding cloud by at least a factor 
of 2.  In contrast with the outflows from other Class 0 objects, the 
CH$_3$OH($2_k$--$1_k$) lines in L483 do not show strong high velocity wings, 
although there is a small ($\sim 0.3$ km s$^{-1}$) velocity shift
approximately along the outflow direction. We do not find 
evidence for a CH$_3$OH abundance enhancement in the flow, and the 
CH$_3$OH lines trace a centrally concentrated core which we model, 
assuming optically thin emission, as having a density gradient
between $r^{-1}$ and $r^{-1.5}$ for radii between $15''$ and $100''$.
H$_2$CO(2$_{12}$--1$_{11}$) lines
show strong high-velocity wings with the same distribution as the outflow,
and evidence for a  H$_2$CO abundance enhancement of a factor of 20 with 
respect to the ambient cloud. At ambient velocities and over the central 
$40''$, this line presents a strong self absorption and 
a brighter blue peak, a characteristic signature of inward motions.
A simple analysis of the H$_2$CO line profiles suggests an infall 
rate of $2\times 10^{-6}$~M$_\odot$ yr$^{-1}$.

Combining the results from our observations with previous work, we 
discuss the evolutionary status of IRAS 18148-0440 and its outflow.
The bipolar outflow presents some characteristics
common to other outflows from Class 0 sources, like high degree of collimation, 
gas heating, and H$_2$CO abundance enhancement. However, other characteristics,
like its low velocity, the lack of bright SiO or CH$_3$OH outflow wings, and
the association with a NIR scattering nebula (optically invisible) seem more
common to outflows from the more evolved Class I sources. As  IRAS 18148-0440
is a Class 0 object based on its spectral energy distribution, we propose that
it is more evolved than other objects in its class, probably in transition
from Class 0 to Class I.


\end{abstract}

\section{Introduction}

Class 0 objects are the youngest stellar objects known (Andr\'e et al. 
\cite{And93}, \cite{And99}). They commonly
power bipolar outflows with extreme characteristics like a very high
degree of collimation and evidence for shock processing of molecular 
gas even in cases of very low stellar luminosity (see Bachiller \& Tafalla 
\cite{Bac99} for a recent review). In order for these outflows
to evolve into the more quiescent (``standard'') 
outflows associated with Class I sources, rapid changes in outflow 
morphology and kinematics have to occur in the few $10^4$ yr
that Class 0 lasts (Andr\'e et al. \cite{And93}). These changes are most 
likely associated with changes in the source itself, which is 
undergoing its major phase of assembling via gravitational 
infall (e.g., Bontemps et al. \cite{Bon96}, Mardones et al. \cite{Mar97}).
Understanding how these first evolutionary changes of the stellar 
and outflow life occur
is a major challenge to star formation studies, and it requires
the simultaneous analysis of Class 0 objects, their outflows, and their 
dense gas environments. Here we present a molecular line study of 
the L483 core and its outflow powered by IRAS 18148-0440 (IRAS 18148 
hereafter), a system that we find at the end of its Class 0 stage, 
starting its transition to become a Class I object.

The source IRAS 18148 in L483, first identified as an embedded object by 
Parker (\cite{Par88a}), is one of the reddest low-mass sources known (Ladd et
al. \cite{Lad91a}, \cite{Lad91b}), and is located toward the Aquila Rift,
at a most likely distance of 200 pc (Dame \& Thaddeus \cite{Dam85}).
Ladd et al. (\cite{Lad91a}) and 
Fuller et al. (\cite{Ful95}) estimate a source bolometric temperature 
(in the sense of Myers \& Ladd \cite{Mye93}) of 50-60 K, and using the 
flux compilation by Fuller et al. (\cite{Ful95}) (their Fig. 4), we
estimate a $L_\mathrm{smm}$/$L_\mathrm{bol} \gtrsim 0.9$~\%
(also, Fuller et al. \cite{Ful95} fit the spectral energy distribution 
with a single-temperature dust model at 40~K, and from their 1.1mm
flux, we estimate $L_\mathrm{bol}$/$L_\mathrm{1.1mm} < 2.5 \times 10^4$). 
These numbers suggest that IRAS 18148 is a Class 0 object
(Andr\'e et al. \cite{And93}, \cite{And99}, Chen et al. \cite{Che95}),
as already proposed by Fuller et al. (\cite{Ful95}) and Fuller \&
Wooten (\cite{Ful00}), although it is less extreme than the 
prototype Class 0 source \object{VLA1623} (Andr\'e et al. \cite{And90},
Andr\'e et al. \cite{And93}).
IRAS 18148  has a luminosity of about 10~L$_\odot$ and
drives a well-collimated bipolar CO outflow (Parker et al. \cite{Par88b}, 
\cite{Par91}, Fuller et al.  \cite{Ful95}, Bontemps et al. \cite{Bon96}, 
Hatchell et al. \cite{Hat99}), and is associated with a variable 
H$_2$O maser (Xiang 
\& Turner \cite{Xia95}) and shocked H$_2$ emission (Fuller et al. \cite{Ful95},
Buckle et al. \cite{Buc99}). NIR imaging of the source vicinity shows a 
well-defined, parabolic reflection nebula, which is optically invisible
and coincides with the blue lobe of the CO outflow 
(Hodapp \cite{Hod94}, Fuller et al. \cite{Ful95}). Ammonia observations
by Goodman et al. (\cite{Goo93}), Fuller \& Myers (\cite{Ful93}), 
Anglada et al. (\cite{Ang97}), and Fuller \& Wootten (\cite{Ful00})
reveal that the L483 core is 
centrally concentrated, has a strong velocity gradient across
it, and a gas kinetic temperature of about 10~K. H$_2$CO and CS 
spectra toward the central 
source present strong self absorption with lines having
brighter blue peak, a signature of infall motions (Myers et al. \cite{Mye95},
Mardones et al. \cite{Mar97}). 

The combination in L483 of Class 0 characteristics, like
a low $T_\mathrm{bol}$ and infall asymmetry, together with the presence of a 
bright NIR nebula, indicative of partial core disruption,
makes this source an interesting object to study the early evolution
of a very young stellar object. To carry out such a study, we have observed
L483 in tracers sensitive to different aspects early stellar life, like
the outflow (CO, section 3.1), the dense core and possible chemical 
outflow anomalies (CH$_3$OH, section 3.2), and infall and shock 
chemistry (H$_2$CO, 3.3). From the combination of these 
observations, we propose that the central source in L483 has already 
started its transition toward Class I, and that the outflow has lost
part of the chemical richness characteristic of Class 0 flows (section 4).

\smallskip
\section{Observations}

We observed L483 with the IRAM 30m telescope during several
sessions in 1994 September, 1995 September, November, and 1996 June. 
Different receiver configurations were used to map the core in
$^{12}$CO(2--1) [230.53799 GHz], H$_2$CO(2$_{12}$--1$_{11}$)
[140.839518 GHz], and CH$_3$OH($2_k$--$1_k$) [96.741420 GHz],
and to observe selected positions in $^{13}$CO(1--0), C$^{18}$O(1--0),
C$^{17}$O(1--0), H$_2^{13}$CO(2$_{12}$--1$_{11}$), 
H$_2$CO(3$_{12}$--2$_{11}$), and SiO(2--1).
Most observations were done in position switching mode (PSW) in order to
obtain flat baselines. After searching for a 
clean off position, we settled with ($-600''$, $300$) with respect to our 
map center ($\alpha_{1950}=18^{\rm h}14^{\rm m}50\fs6$, 
$\delta_{1950}=-4\degr40'49\farcs0,$
position of IRAS 18148-0440). This position lies outside the optical 
obscuration associated with L483 and seems free from 
dense gas tracer emission, although it has a weak $^{12}$CO(2--1)
line at a level of 2~K between $V_\mathrm{LSR}=6-9$ km s$^{-1}$
(outside the L483 range). A frequency switched (FSW) $^{12}$CO(2--1) 
spectrum of this 
position was taken so it could be added to the data if needed, and a 
test was made by adding the off spectrum to a PSW
spectrum from the origin and comparing the result 
with a FSW spectrum; 
the two were indistinguishable. The data shown in this
paper, except when indicated, correspond to PSW observations without
addition of the off position. 

The backend was an autocorrelator split into different 
windows with resolutions ranging from 0.1 km s$^{-1}$ for $^{12}$CO(2--1)
to 0.03 km s$^{-1}$ for CH$_3$OH($2_k$--$1_k$). The telescope pointing 
was corrected frequently by observing continuum sources and is expected to be
accurate within $3''$. The $T_\mathrm{A}^*$ scale of the telescope was 
converted into $T_\mathrm{mb}$ using the main beam efficiencies recommended 
by Wild (\cite{Wil95}). The full width at half maximum of the 
telescope beam varies linearly with wavelength, and ranges from $25''$ for 
CH$_3$OH($2_k$--$1_k$) to $11''$ for $^{12}$CO(2--1). Spectra were taken with
$10''$ spacing, which is slightly less than one beam at the highest frequency.

\section{Results}

\subsection{CO data}

\subsubsection{Ambient CO emission} 

\begin{figure}
\resizebox{\hsize}{!}{\includegraphics{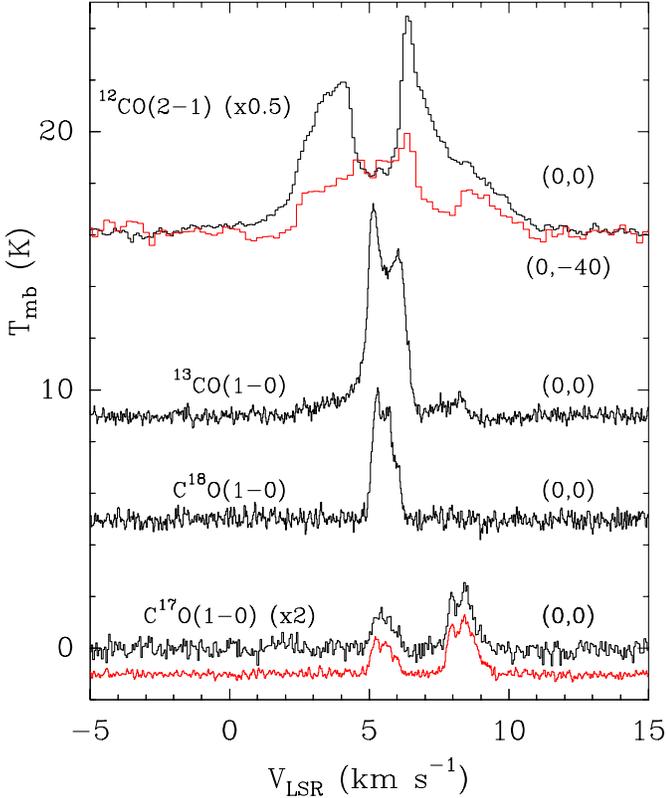}}
\caption{Spectra of CO isotopomers toward IRAS 18148-0440, at the center
of the L483 core ($\alpha_{1950}=18^{\rm h}14^{\rm m}50\fs6$,
$\delta_{1950}=-4\degr40'49\farcs0$). For $^{12}$CO(2--1), the
spectrum toward the non outflow position (0, $-40''$) is also shown in
lighter shade to better illustrate the deep self absorption and
outflow wings toward the central position.
Both $^{12}$CO(2--1) spectra have been corrected for emission in the
off position. Below the C$^{17}$O(1--0) spectrum, and in lighter shade,
is shown a model using three replicas of the C$^{18}$O(1--0)
line with the relative weights expected for optically thin emission. Note
the good agreement, which indicates a negligible optical depth in
C$^{17}$O(1--0).
}
\label{fig1}
\end{figure}          

In this section we derive the basic parameters of the 
gas along the line of sight towards the L483 core center
using a series of CO isotopomer lines.
As Fig.~\ref{fig1} shows, the $^{12}$CO(2--1) line is heavily self absorbed
between $V_{LSR}$  4.5 and 6.5 km s$^{-1}$, with a $T_\mathrm{mb}$ of 4-5~K
Given that the gas between these velocities is optically thick, its 
excitation
temperature $T_\mathrm{ex}=$ has to be 8.5-9.5~K, a value
close to the kinetic temperature of the dense gas (10~K, from  NH$_3$ 
and HC$_3$N, Fuller \& Myers \cite{Ful93}, Anglada et al.
\cite{Ang97}). The gas 
responsible for the $^{12}$CO(2--1) absorption represents low
density gas in the outer layers of the cloud, far from the dense 
core, so the above result suggests that all the ambient gas along the line 
of sight has an almost constant temperature of 10~K (note that 
the outflow gas is warmer, see below). 

In contrast with the self absorbed $^{12}$CO(2--1) emission, 
C$^{18}$O(1--0) and C$^{17}$O(1--0) are optically thin.
This is not evident because of the non-Gaussian shape of the spectra 
(Fig.~\ref{fig1}), but can be proved using the 
following simple model: we create a C$^{17}$O(1--0) spectrum by adding three
replicas of the C$^{18}$O(1--0) line each shifted in velocity
by the proper hyperfine splitting and weighted by the optically
thin relative intensity. The result, shown in lighter
shade below the C$^{17}$O(1--0) spectrum, matches very well the
observations, indicating  that both C$^{18}$O(1--0) and C$^{17}$O(1--0)
are thin. The non Gaussian shape of these lines has therefore 
to result from velocity structure along the line of sight
(see section 3.2 for further details).

With the excitation temperature of 10~K and the fact
that C$^{17}$O(1--0) is thin, we can derive the core central
H$_2$ column density. We integrate the C$^{17}$O(1--0) emission in velocity
and assume local thermodynamic equilibrium, estimating
a N(C$^{17}$O) of 1.5 $\times 10^{15}$ cm$^{-2}$. 
For a standard C$^{17}$O abundance of
$4.7 \times 10^{-8}$ (Frerking et al. \cite{Fre82}, Wilson \& Rood
\cite{Wil94}), this value implies an H$_2$ column density of
3 $\times 10^{22}$ cm$^{-2}$.

To finish this section, we note
the presence in Fig.~\ref{fig1} of CO emission outside the ambient cloud 
range ($V_{LSR}= 4.5$-6.5 km s$^{-1}$). Part of this
emission comes from outflow gas (discussed in section 3.1.2),
but other part must come from additional clouds along the line of sight.
This is the case for two features at $V_{LSR}=3$ and 8 km s$^{-1}$, 
mostly seen in $^{13}$CO, as they are spread over an area larger than 6 
arcminutes.
It must also be the case for some contribution between 2.5 and 4.5 km s$^{-1}$,
which is very prominent in the $^{12}$CO spectra outside the outflow
range (see (0,$-40$) spectrum in Fig.~\ref{fig1}). As we will see
below, this low velocity emission limits our study of the outflow gas,
to which we now turn our attention.

\subsubsection{The outflow}

\begin{figure*}
  \resizebox{\hsize}{!}{\includegraphics{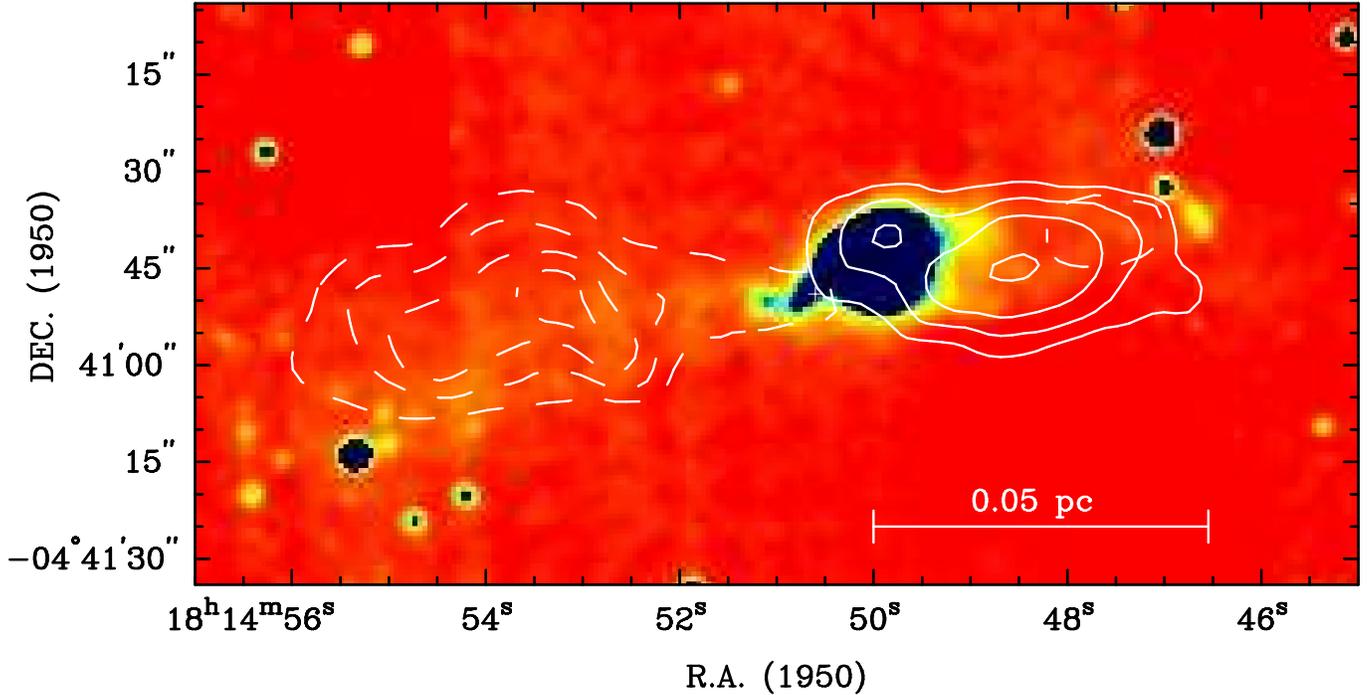}}
  \caption{High velocity $^{12}$CO(2--1) emission superposed to
the K$'$ (2.2~$\mu$m) image of Hodapp (\cite{Hod94}) illustrating the 
excellent agreement
between the outflow distribution and the scattering nebula. 
Solid contours represent blue emission integrated from $V_\mathrm{LSR}$ 0 to 
2 km s$^{-1}$ and dashed contours represent red emission 
integrated from $V_\mathrm{LSR}$ 8 to 10 km s$^{-1}$.
For both lobes, contours are at 3, 6, ... K km s$^{-1}$. 
Astrometry for the K$'$ image has been done by finding common 
stars in the DSS image, and the estimated positional accuracy is less than 
$2''$. The K$'$ image has been convolved with a Gaussian beam 
to enhance the diffuse emission from the red lobe. The
star sign marks the position of IRAS 18148 and the scale bar is for an 
assumed distance of 200 pc.
 }
\label{fig2}
\end{figure*}

Fig.~\ref{fig2} presents our CO(2--1) map of the L483 outflow
(see also Fig. ~\ref{app_fig2}). 
As previous maps (Parker et al.
\cite{Par91}, Fuller et al. \cite{Ful95}, Bontemps et al. (\cite{Bon96}),
and Hatchell et al. \cite{Hat99}), it shows that
the accelerated CO emission extends E-W with 
IRAS 18148 at the center. The red CO lies to the east of IRAS 18148 while
the blue CO lies to the west, although there is
some red gas near the western tip of the blue lobe. This anomalous red gas
coincides with a region of shocked H$_2$ emission (Fuller et al. 
\cite{Ful95}), and probably represents outflow material with an 
enhanced turbulent component because of the shock.

A comparison of the CO outflow with the K$'$ (2.1 $\mu$m)
image of Hodapp (\cite{Hod94}) (Fig.~\ref{fig2}) shows that
the blue CO delineates very closely the reflection
nebula and presents a relative maximum toward its center. In contrast, 
the red (eastern) CO is weak and highly
collimated near IRAS 18148 and reaches a maximum 
0.04~pc away from it. This east-west asymmetry of the CO
outflow suggests that a similar asymmetry
occurs in the underlying nebula, which would
therefore have an intrinsically brighter western side.

As the $^{12}$CO(2--1) spectrum in Fig.~\ref{fig1} shows, 
the outflow gas is warmer than the ambient core, or otherwise
its emission would not be brighter than the self absorption
by 10~K gas.
The strongest CO wings imply excitation temperatures of 20~K 
(twice the ambient kinetic temperature), and this is a lower limit 
because the CO emission may not be thick and thermalized.
In fact, Hatchell et al. (\cite{Hat99}) have argued, from a comparison of
CO(4--3) and CO(2--1) lines, that temperatures up to 50~K could be present
in the CO outflow. Given that we do not find brightness temperatures
larger than 20~K, we take this value as a lower limit  for our further 
calculations of the outflow energetics (see below) and molecular
abundances (sections 3.2 and 3.3).

With the above value for the CO(2--1) $T_{ex}$, we estimate the 
outflow energetics. 
This requires special care because the outflow is rather slow and
the CO spectra are contaminated at low velocities, so
in the Appendix we present a method to deal with the contaminating
{\em emission}.  Of course, we cannot correct for the
outflow gas hidden by the self absorption,
so all CO emission in the central 3 km s$^{-1}$
(about $V_\mathrm{LSR} = 5.5$~km s$^{-1}$) will be
ignored, and our estimate will represent a lower limit. As the highest 
outflow velocities, we take $V_{LSR}$ $-3$ and 14~km s$^{-1}$, because no
faster outflow emission is found in the CO(2--1) velocity maps.
Assuming a CO abundance of $8.5\times 10^{-5}$
(Frerking et al. \cite{Fre82}), we 
derive an outflow mass of 0.01 M$_\odot$,
a momentum of 0.03~M$_\odot$ km s$^{-1}$, and a kinetic energy of
$2 \times 10^{42}$ erg. From a total outflow length of $160''$ and
a total velocity extent of 15~km s$^{-1}$, we derive a kinematical
time of $10^4$~yr. These values are in reasonable agreement with
those from from Parker et al. (\cite{Par91}). 
 
\subsection{CH$_3$OH and SiO data}

\begin{figure}
  \resizebox{\hsize}{!}{\includegraphics{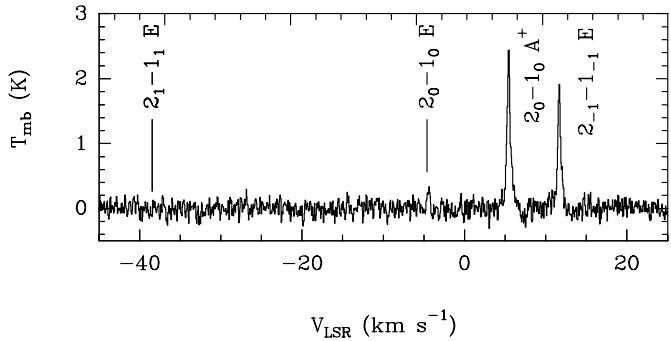}}
  \caption{CH$_3$OH(2$_k$--1$_k$) spectrum toward the center of the L483
core showing the different $k$ components (see Fig.~\ref{fig1} for 
coordinates).
}
\label{fig3}
\end{figure}

	Figure~\ref{fig3} presents a CH$_3$OH($2_k$--$1_k$) spectrum 
toward the core center with labels indicating the different $k$ components.
Among the E-type lines, the $k$=1 component is not detected, while the
the integrated intensity ratio between the $2_{-1}$--$1_{-1}$
and $2_0$--$1_0$ lines is approximately 7, very similar to the ratio found
by Turner (\cite{Tur98}) for \object{TMC-1} and \object{L183}. 
The relatively high intensity
of the lines (about 2.5~K at the peak) implies an excitation temperature
of at least 5.5~K, while an LTE rotation diagram analysis (e.g., Menten 
et al. \cite{Men88}) indicates temperatures of 4~K at most. This suggests that
non LTE conditions may apply, and that for the E-type CH$_3$OH, the
lower lying $k$=$-1$ ladder has a higher $T_\mathrm{ex}$ than the 
higher $k$=0 ladder. The excitation conditions are
probably constant over the core, as an average over all
(75) positions outside the central $20''$ gives a spectrum with the
same intensity ratios among all the E and A components as in the spectrum shown
in Fig.~\ref{fig3}; we will use below this apparently constant conditions to
infer a lower limit to the density gradient in the core. From the integrated
intensities in the central spectrum, we estimate a CH$_3$OH column density of 
about $7 \times 10^{13}$~cm$^{-2}$, which together with our estimated
H$_2$ column density implies a CH$_3$OH abundance of $2 \times10^{-9}$,
which is very close to the abundances estimated for other dark clouds
(Friberg et al. \cite{Fri88}, Bachiller et al. \cite{Bac95}, Turner 
\cite{Tur98}). 

\begin{figure*}
  \resizebox{\hsize}{!}{\includegraphics{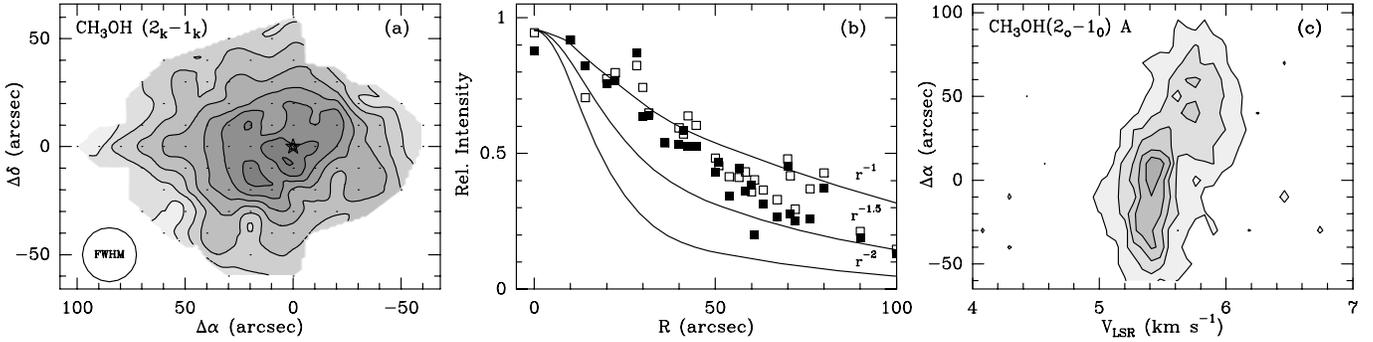}}
  \caption{{\bf a} Map of CH$_3$OH(2$_k$--1$_k$) intensity 
integrated over the $2_{0}$--$1_{0}$ A and $2_{-1}$--$1_{-1}$ E components.
Levels go from 0.3 to 2.7 K km s$^{-1}$ in steps of 0.3 K km s$^{-1}$.
{\bf b} Radial average of integrated intensity (filled squares represent 
$2_{-1}$--$1_{-1}$ E and open squares represent $2_{0}$--$1_{0}$ A)
with optically thin, LTE models for different density laws
convolved with a $25''$ Gaussian (solid lines).
{\bf c} Position-velocity diagram along R.A. for $\Delta\delta=0$
(i.e., along the outflow axis).
Contours go from 0.5 to 2.5~K in steps of 0.5~K. Note the velocity 
shift near $\Delta\alpha=30''$. Offsets with respect to IRAS 18148
(see Fig.~\ref{fig1})
}
\label{fig4}
\end{figure*}

The spatial distribution of the CH$_3$OH($2_k$--$1_k$) integrated intensity, 
shown in  Fig.~\ref{fig4}a, is rather round and centrally concentrated 
toward IRAS 18148. This is also the case with other dense gas tracers 
like HC$_3$N, NH$_3$, and N$_2$H$^+$ (Fuller \& Myers \cite{Ful93}, 
Anglada et al. \cite{Ang97}, Caselli et al. \cite{Cas00}), and is probably a 
sign of the extreme youth of the embedded star, which has not had
time to perturb the bulk of the parental core (despite carving the nebula).
The central concentration of the emission
reinforces our interpretation that the emission is
mostly optically thin, as otherwise we would expect a flat distribution
(our unpublished NH$_3$ data indicate a constant gas kinetic
temperature across the core).
 
Fuller et al. (\cite{Ful95}) have inferred an r$^{-2}$ density gradient
within $30''$ of the IRAS source based on the intensity contrast between 
the two sides
of the IR reflection nebula, but their calculation depends critically
on assuming that the nebula is symmetric, which we have seen is probably not 
the case. More recently, Fuller \& Wootten (2000) have proposed that 
the r$^{-2}$ profile continues to larger radii ($100''$) using a model 
for the NH$_3(1,1)$ emission, although this result could be 
affected by interferometer missing flux.
Here we use the CH$_3$OH emission to derive an independent estimate,
assuming that this emission is thin and the excitation temperature
is constant. In Fig.~\ref{fig4}b, we show a radial average of the emission, 
together with the results from our model for three density power
laws: $r^{-1}$, $r^{-1.5}$, and $r^{-2}$ (results convolved 
with a $25''$ Gaussian beam). As the figure shows, the $r^{-2}$ density
law is too steep, and the observations are better fit between $R = 15''$
and $100''$ using a $r^{-1}$ profile with probably some steepening 
for $R > 50''$.  

In the optically thin limit, the brightness radial profiles 
of Fig.~\ref{fig4} are proportional 
to column density radial profiles, so we can use them to estimate the core 
mass by normalizing them to the central H$_2$ column density and
integrating them radially. In this way, we derive a core mass between
5~M$_\odot$ ($r^{-1.5}$ density profile) and 10~M$_\odot$ ($r^{-1}$ density
profile). These values are in good agreement with the ammonia
result from Anglada et al. \cite{Ang97}, and also agree with a virial
estimate using the average CH$_3$OH line width over the core 
(0.64~km s$^{-1}$), which gives 7 and 8~M$_\odot$ for $r^{-1.5}$ and
$r^{-1}$ density profiles, respectively. 

Although the CH$_3$OH lines are relatively narrow compared with other
dense gas tracers like H$_2$CO, they are systematically asymmetric and
change velocity across the core. This is illustrated in Fig.~\ref{fig4}c  
with a position-velocity diagram along the east-west axis of the core, 
which shows that the line center velocity
shifts abruptly by about 0.3 km s$^{-1}$
near the position of IRAS 18148 (origin of
offset coordinates); lines toward  
$\Delta\alpha \le 20''$ are brighter, bluer, and narrower than toward 
the east. These two velocity components coincide with the
C$^{18}$O (and C$^{17}$O) peaks we have found in section 3.1.1,
and and can also be seen in our N$_2$H$^+$(1--0) data.
Fuller \& Myers (\cite{Ful93}),  
with low resolution observations, have reported
that the spectra from L483 seem to have two velocity components, while 
Goodman et al. \cite{Goo93} found a systematic velocity gradient across the 
core, all in the same direction as the velocity change we find.
The origin of this behavior is not clear, but we notice that
the sense of the velocity change agrees with the sense of the bipolar
outflow. It
is therefore possible that it arises from the acceleration of 
dense gas by the outflow, since the position velocity diagram
is similar to that of \object{L1228} in C$_3$H$_2$ (Tafalla \& Myers 
\cite{Taf97}),
where outflow acceleration is the cause of a similar velocity shift.

Independently of the origin of the velocity shift, it is clear that
the CH$_3$OH spectra in L483 do not show the prominent high-velocity 
wings seen in some 
low mass outflows like \object{NGC 1333-IRAS 2} (Sandell et al. \cite{San94})
and L1157 (Bachiller at al. \cite{Bac95}). The wings in these systems 
arise from large abundance enhancements of CH$_3$OH, which in the best studied
case of L1157, amounts to a factor of 400 (Bachiller et al. \cite{Bac95}, 
Bachiller \& P\'erez-Guti\'errez \cite{Bac97}). For L483, the non detection
of CH$_3$OH at the velocities with CO outflow implies
that any possible CH$_3$OH abundance enhancement cannot be larger than
around 10, which is at least 40 times lower than in L1157. Higher 
signal-to-noise
data can probably lower this limit by a significant amount. 

Another tracer with prominent wings toward certain outflows is
SiO (e.g., Bachiller \& P\'erez-Guti\'errez \cite{Bac97}), and three
L483 positions were observed in the 2--1 line of this molecule (origin, 
$30''$ E, and $30''$ W). Our non detections (with limits of the 
order of 0.1 K km s$^{-1}$) imply SiO abundances lower than
$10^{-11}$ and $8 \times 10^{-10}$ for the ambient gas and
the outflow, respectively (numbers derived using an large velocity
gradient analysis). Although the non detections make impossible
to estimate an abundance enhancement in the outflow, 
comparing our outflow limit with the SiO abundance found
in L1157 ($10^{-7}$, Mikami et al. \cite{Mik92}, Bachiller 
\& P\'erez-Guti\'errez \cite{Bac97}), we estimate that any
possible enhancement in L483 is at least 100 times smaller
than in L1157.

\subsection{H$_2$CO Data}

\begin{figure}
  \resizebox{\hsize}{!}{\includegraphics{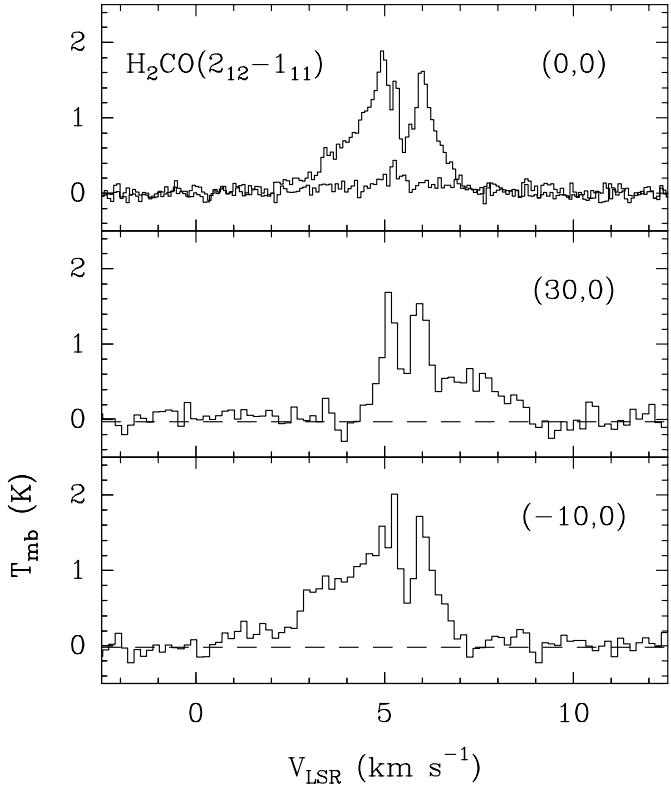}}
  \caption{H$_2$CO(2$_{12}$--1$_{11}$) spectra toward selected positions 
of the L483 core. For the top panel, the weaker 
H$_2^{13}$CO(2$_{12}$--1$_{11}$) is also shown.
Offset are in arcsec with respect to the position of IRAS 18148 
(see Fig. \ref{fig1}).
}
\label{fig5}
\end{figure}

Our other dense gas tracer, H$_2$CO, presents very different lines than 
CH$_3$OH. Fig.~\ref{fig5} shows a series of
H$_2$CO(2$_{12}$--1$_{11}$) spectra for different core positions
illustrating the variety of profiles. 
At ambient velocities ($V_\mathrm{LSR}\sim5.5$~km/s), the 
lines are strongly self absorbed over most of the core, as 
can be checked by comparing the double-peaked 
main isotope with the single-peaked rare H$_2^{13}$CO line
toward the central position (Fig.~\ref{fig5} top panel). 
This self absorption is slightly
red shifted with respect to the emitting gas, so the resulting spectrum 
has a brighter blue peak, a fact already noticed
by Myers et al. (\cite{Mye95}) and Mardones et al. (\cite{Mar97}), who have
presented preliminary versions of the H$_2$CO spectrum towards the core 
center. Such red-shifted self absorptions are characteristic signatures 
of inward motions (e.g, Leung \& Brown \cite{Leu77}), and their presence 
in L483 makes this core one of the best infall candidates known. 
A detailed study of the spatial distribution of
the self absorption and its interpretation in terms of infall motions will
be presented elsewhere (Mardones et al. \cite{Mar00}, see also Mardones 
\cite{Mar98}), so
here we limit ourselves to briefly comment on this feature.

Double-peaked H$_2$CO spectra abound in the core and are more prominent 
along the outflow axis, although a map of the spectral asymmetry parameter
$\delta v$ (defined as the difference between the thick and thin line peaks
normalized to the thin line width, see Mardones et al. \cite{Mar97}) shows that 
the blue asymmetry is stronger perpendicular to the flow. Overall, 
despite the presence of high-velocity blue and red wings in the 
H$_2$CO lines, the ambient self absorption is mostly red shifted and an 
average spectrum over the central core is clearly asymmetric in the sense 
of infall. To estimate the global infall rate in L483, we first determine the
infall radius from the extent of the H$_2$CO spectra with brighter blue peak,
which we measure from the data as 0.02~pc. Then, we use
the simple 2-layer model of Myers et al. (\cite{Mye96}), which allows to
derive an infall velocity from the contrast between the blue and 
red peaks of a self absorbed line profile knowing the intrinsic line width 
from a thin tracer (see their Eq. 9). To do this, we take 
the average H$_2$CO spectrum inside the infall radius, and use as 
thin tracer the average CH$_3$OH spectrum over the same area. In this
way, we derive a mean infall speed of 0.02~km s$^{-1}$, which is clearly 
subsonic. Finally, we derive a mean density inside the infall radius
using the power-law density models derived in our CH$_3$OH analysis,
which give a mean density of $3.3\times 10^5$
cm$^{-3}$ for a $r^{-1}$ profile and $2.8\times 10^5$ 
cm$^{-3}$ for a $r^{-1.5}$ profile. Averaging the above densities
to $3\times 10^5$ cm$^{-3}$, we derive a mass infall rate of 
$2\times 10^{-6}$~M$_\odot$ yr$^{-1}$ (see Mardones et al. \cite{Mar00} for 
further details).

\begin{figure*}
  \resizebox{\hsize}{!}{\includegraphics{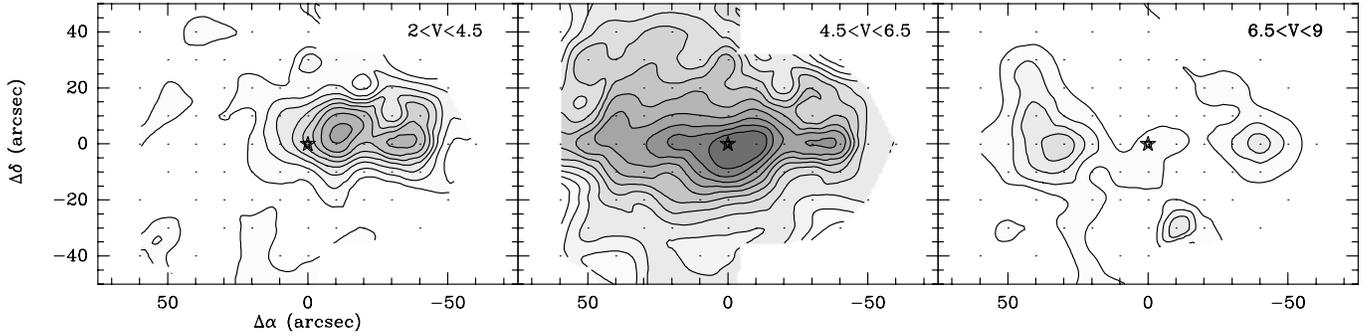}}
  \caption{H$_2$CO(2$_{12}$--1$_{11}$) velocity maps showing blue, ambient,
and red emission (left, center, and right panels). Note how the H$_2$CO
distribution follows that of the CO outflow (Fig.~\ref{fig2}). 
In all panels, the the first level and the level
step is 0.2 K km s$^{-1}$. Offsets as in Fig. 4a.
        }
\label{fig6}
\end{figure*}

At velocities outside the ambient range, the H$_2$CO spectra present strong 
wings that change with position following the
distribution of accelerated gas in the bipolar outflow. The  
spectra at ($30''$, 0) and ($-10''$, 0) in Fig.~\ref{fig5} illustrate this
effect, which implies that the bipolar outflow is accelerating part of the 
dense core gas, probably shocking it and altering
its chemical composition (see below). 
To better illustrate the dense-gas acceleration traced by H$_2$CO, we
present in Figure~\ref{fig6} velocity maps for the blue, ambient,
and red regimes. The blue and red maps agree very well with equivalent
outflow maps from CO (see Fig.~\ref{fig2}), especially for the blue lobe.
There, both emissions present two maxima, one associated with the reflection
nebula, and the other with the region of strong H$_2$ emission at the end
of the lobe. Even more, both emissions present an ``anomalous'' red peak towards
the position of bright H$_2$ emission, again suggesting this gas 
is shock related. These
similarities suggest that CO and H$_2$CO, despite their different 
dipole moments, are tracing the same (or very closely connected) gas. 

\begin{figure}
  \resizebox{\hsize}{!}{\includegraphics{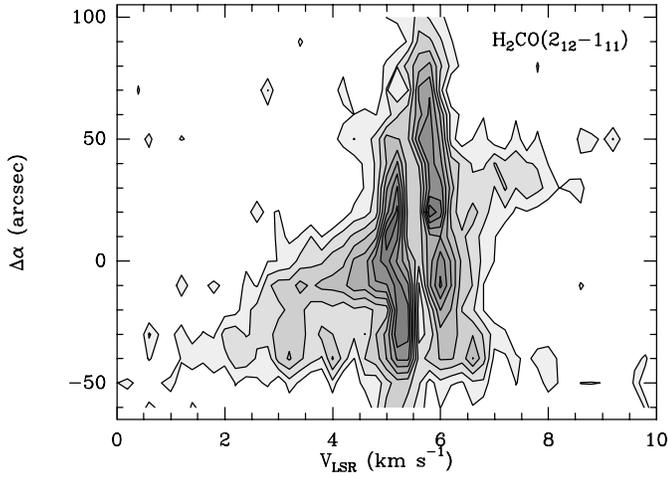}}
  \caption{Position-velocity diagram of the H$_2$CO(2$_{12}$--1$_{11}$)
emission along R.A. for $\Delta\delta=0$ (i.e., along the outflow axis).
Note the prominent outflow wings increasing in velocity with position;
compare with Fig. 4c.  First level and step are 0.2~K.
        }
\label{fig7}
\end{figure}

As a further illustration of the importance of the outflow acceleration in 
the H$_2$CO emission, we show in Fig.~\ref{fig7} a position-velocity 
diagram along the the Right Ascension axis (i.e., approximately
parallel to the outflow). At ambient velocities, the
strong self absorption can be easily seen in the form of two 
bright emission peaks at about 5.2 and 6 km s$^{-1}$. At high
velocities, the outflow wings form two triangular-shaped extensions
which illustrate how the emission terminal velocity increases almost
linearly with distance from the outflow source (at $\Delta\alpha = 0$);
such linear increases are common in the CO emission from outflows
(e.g., Meyers-Rice \& Lada \cite{Mey91}). 
We can also see in the position-velocity diagram that towards the
west ($\Delta\alpha \sim -50''$), both the wing and the ambient emission
drop simultaneously. This suggests that the high velocity 
gas is breaking through the dense core, something also suggested 
by the presence of bright H$_2$ emission.
Towards the east, the ambient gas is more extended than the 
outflow wing, so it seems the outflow is bounded in this
direction.

We finish studying the H$_2$CO abundance in the outflow and comparing it 
to the abundance in the ambient cloud. As the 
H$_2$CO(2$_{12}$--1$_{11}$) line is self absorbed at ambient velocities,
we use in this range the H$_2^{13}$CO isotopomer, which is very likely
thin (see Fig.~\ref{fig5}).  In this way, using an LTE analysis with 
$T_\mathrm{ex}=5-10$~K
and a $^{12}$C/$^{13}$C ratio of 77 (Wilson \& Rood \cite{Wil94}),
we derive a H$_2$CO ambient column density of $5 \times 10^{13}$
cm$^{-2}$ for the central position (a similar analysis using 
H$_2$CO(3$_{12}$--2$_{11}$) gives the same number).
This value implies an ambient abundance of $1.5 \times 10^{-9}$.
For the outflow regime, we use the H$_2$CO(2$_{12}$--1$_{11}$)
line and compare it with the outflow CO emission.
Applying the same LTE analysis as before, we estimate outflow
abundances at ($-10''$, 0) (blue) and ($30''$, 0) (red) of about 
$3\times 10^{-8}$, suggesting an H$_2$CO abundance enhancement 
of a factor of 20. This number is slightly smaller that the factor of
60-80 found in L1157 (Bachiller \& P\'erez-Guti\'errez \cite{Bac97}),
but is significantly larger than our limit for the CH$_3$OH enhancement.
Thus, in contrast with L1157 for which CH$_3$OH is enhanced by almost 
an order of magnitude more than H$_2$CO, L483 is richer in
H$_2$CO.

\section{Evolutionary status of IRAS 18148-0440}

As mentioned in the Introduction, 
IRAS 18148 belongs to Class 0 (Andr\'e et al. \cite{And93}, 
Andr\'e et al.  \cite{And99}) due to its red 
spectral energy distribution, although it is less extreme 
than objects like VLA1623. IRAS 18148 has additional
Class 0 characteristics like the presence of infall asymmetry in 
line spectra (Mardones et al. \cite{Mar97}). 
After analyzing its molecular environment in previous sections, 
we now compare this object and its outflow with other Class 0
sources, and to do this, we present in Table 1 a
summary of properties of different Class 0 sources
with similar luminosity ($\sim 10 L_\odot$)
for which enough molecular data are available.

\begin{table}
   \caption[]{Class 0 sources and their outflows}
       \label{tab1}
    \[
         \begin{array}{lrrrccccc}
            \hline
            \noalign{\smallskip}
              &  & & 
            \multicolumn{2}{c}{\mbox{Outflow}} & 
            \multicolumn{3}{c}{\mbox{Chem. Enhancement$^{\mathrm{a}}$}} & 
             \\
             \mbox{Name} & L_{bol} & T_{bol} & V_{max}^{\mathrm{b}} & 
             \mbox{B}^{\mathrm{c}} &
             \mbox{SiO} & \mbox{CH$_3$OH} & \mbox{H$_2$CO} & \mbox{Ref} \\
              & \mbox{L$_\odot$} & \mbox{K} &  & & & & & \\
            \noalign{\smallskip}
            \hline
            \noalign{\smallskip}
                \mbox{L1448-C} & 9 & 56 & 70 & \mbox{Y} & >10^5 & ? & ? & 
                1,2,3,4 \\
                \mbox{L1157} & 11 & 62 & 12 & \mbox{N} & >10^5 & 400 & 75 & 
                2,5,6,7\\
                \mbox{BHR 71} & 9 & 56 & 10 & \mbox{N} & 200 & 30 & ? & 
                2,8,9 \\
                \mbox{L483} & 10 & 50 & 7.5 & \mbox{N} & 
                \mbox{ND$^{\mathrm{d}}$} & <10 & 20 & 2,10\\
            \noalign{\smallskip}
            \hline
         \end{array}
      \]
   \begin{list}{}{}
     \item[$^{\mathrm{a}}$] With respect to ambient cloud.
     \item[$^{\mathrm{b}}$] Outflow terminal velocity from CO in km s$^{-1}$.
     \item[$^{\mathrm{c}}$] Presence of CO ``bullets.''
     \item[$^{\mathrm{d}}$] Not detected. Enhancement at least 100 times
     smaller than in L1157.
     \item[References:] (1) Bachiller et al. \cite{Bac91a}, (2) Mardones
     et al. \cite{Mar97}, (3) Bachiller et al. \cite{Bac90},  (4) Bachiller 
     et al.  \cite{Bac91b}, (5) Umemoto et al. \cite{Ume92}, (6)  Bachiller \&
     P\'erez-Guti\'errez \cite{Bac97}, (7) Avery \& Chiao \cite{Ave96},
     (8) Bourke et al. \cite{Bou97}, (9)
     Garay et al. \cite{Gar98}, (10) Fuller et al. \cite{Ful95}
   \end{list}
\end{table}

As Table 1 shows, L483 is very different 
kinematically from \object{L1448-C}, a
source characterized by an extremely collimated and fast bipolar CO outflow
with discrete high velocity components (``bullets'') (Bachiller et al. 
\cite{Bac90}, see also Barsony et al. \cite{Bar98}, O'Linger et
al. \cite{Oli99}, Eisl\"offel \cite{Eis00}, and 
the similar object  VLA1623 found by
Andr\'e et al. \cite{And90}). Despite our limited knowledge of 
the chemical composition of the L1448-C outflow, Table 1 
shows that L483 also differs from L1448-C in its chemical 
enhancements, again in the sense of L1448-C being more extreme.

More similarities are found between L483 and L1157 and \object{BHR 71}
(L1157: Umemoto et al. \cite{Ume92}, Mikami et al. \cite{Mik92}, 
Bachiller et al. \cite{Bac95}, Tafalla \& Bachiller \cite{Taf95}, 
Zhang et al. \cite{Zha95}, Avery \& Chiao \cite{Ave96}, 
Bachiller \& P\'erez-Guti\'errez \cite{Bac97}
Gueth et al. \cite{Gue98}, Umemoto et al. \cite{Ume99}, Zhang et al. 
\cite{Zha00}; BHR 71: Bourke et al. \cite{Bou97}, Garay et al. \cite{Gar98}). 
These
outflows have comparable kinematics in the sense of having lower
velocities and lacking ``bullets'' (also their collimation is similar),
but differ by the amount of the chemical enhancement in the 
sense that the numbers for L483 are systematically smaller.
Although the uncertainties in the chemical estimates are rather large,
the differences shown in Table 1 are too extreme to be 
due to observational error. This means that among ``chemically active'' 
outflows, L483 is a weak case.

Bachiller \& Tafalla \cite{Bac99} have argued that outflows evolve 
with time from 
having ``bullets'' to not having them, and from being chemically
active to not being so (see Bontemps et al. \cite{Bon96} for
evidence that outflow momentum flux systematically decreases 
with outflow age). If this 
evolution scenario is correct, 
IRAS 18148 in L483 would represent a rather late stage of 
a Class 0 source, and the objects in Table 1 would be ordered by 
increasing age from top to bottom. Unfortunately, 
no independent stellar clock exists yet to order by age 
the different Class 0 objects, and as Table 1 shows, the bolometric
temperature $T_\mathrm{bol}$ cannot distinguish between extremely
young sources, probably due to the lack of mid and far IR 
photometry. There is a reason, however, to suspect that IRAS 18148
is more evolved than other Class 0 sources, and that the above 
evolution scenario is correct: L483 has a bright NIR nebula
(Hodapp \cite{Hod94}, Fuller et al. \cite{Ful95})
in contrast with L1448-C and L1157 (Bally et al. \cite{Bal93},  
Hodapp \cite{Hod94}, Davis \& Eisl\"offel \cite{Dav95})
(but note BHR 71 also has a NIR nebula, Bourke et al. \cite{Bou97}).
The presence of such a nebula suggests that the L483 outflow has been
accelerating ambient material for longer than the outflows
from the other objects in Table 1, and that IRAS 18148 is
more advanced in its transition to become a visible object
(e.g., Shu et al. 1987).
Other Class 0 object with a NIR nebula is L1527 (Eiroa et al. \cite{Eir94}),
so if the above scenario is correct, one would expect
to find in this object a weak abundance enhancement 
of the molecules shown in Table 1. Observations of this object 
should be done to test this point.

The status of L483 as a somewhat evolved Class 0 object is also
consistent with its relatively low value 
$L_\mathrm{smm}$/$L_\mathrm{bol}$ ($\gtrsim 0.9$~\%,
compare with the 10~\% 
of VLA1623, Andr\'e et al. \cite{And93}) and 
with the work of Bontemps et al. \cite{Bon96}. Applying
the factor of 10 correction these authors apply to their CO data 
(see their Eq. 2), we estimate for L483 a momentum flux {\em \'a la}
Bontemps et al. of $3 \times 10^{-5}$~M$_\odot$ km s$^{-1}$~yr$^{-1}$.
This value is almost a factor of 2 lower than the mean momentum flux
of Class 0 objects, but still 8 times larger than the average
number for Class I sources. L483, again, appears as a  Class 0
object already evolving toward Class I.

If the Class 0 encompasses sources as diverse as those powering the
L1448-C and L483 outflows, outflow evolution should occur very rapidly
during the star's first few $10^4$~yr (the expected 
duration of the Class 0 stage, see Andr\'e et al. \cite{And93}). 
It is possible that this evolution is driven by a rapid decrease 
in the infall/accretion rate on the central object
(Bontemps et al. \cite{Bon96}, Tomisaka \cite{Tom96}, Henriksen et al. 
\cite{Hen97}), but although
Class 0 objects do have stronger infall signatures 
than Class I sources, there is no clear infall trend among
Class 0 objects themselves (Mardones et al. \cite{Mar97}).
Further study of transition objects like L483 is needed to
understand these earliest changes of stellar life, and as
this work has shown, the combination of the chemistry and 
kinematics of the outflow may hold the key to that understanding.

\section{Summary}

We have observed the L483 core and outflow in different mm molecular
transitions and made full maps in CO(2--1), CH$_3$OH(2$_k$--1$_k$), 
and H$_2$CO(2$_{21}$--1$_{11}$). With these data, 
we have studied the outflow, the core, and their relation with 
the IR cometary nebula around IRAS 18148. The
main conclusions of our work are as follows:

1. The $^{12}$CO emission at ambient velocities is extremely 
thick with the brightness temperature expected for gas at 9~K, 
the temperature previously estimated for the core gas. Outside
the ambient regime the $^{12}$CO lines present bright wings 
indicating outflow material warmer than the ambient gas by at least 
a factor of 2. A simple model for the  C$^{17}$O(1--0) emission 
towards the core center shows that this line
is optically thin and non Gaussian due to the presence
of two velocity components. From the integrated C$^{17}$O(1--0) emission
we estimate a central H$_2$ column density of $3 \times 10^{22}$ cm$^{-3}$
in the inner $20''$.

2. The CO outflow emission is compact and slow, with a total
length of 0.15 pc and a kinematical age of $10^4$~yr. Lower limits
to the outflow mass, momentum, and energy are 0.01~M$_\odot$, 0.03~M$_\odot$ 
km s$^{-1}$, and $2\times 10^{42}$ erg, respectively. The CO outflow 
is asymmetric, with a blue lobe having a bright spot coinciding with 
the NIR nebula and the red lobe being weaker near IRAS 18148
and having a relative maximum $45''$ from the source. This
asymmetry suggests that the reflection nebula around the 
IRAS source may also be asymmetric and have a more prominent blue side.

3. The CH$_3$OH emission traces a dense core with no appreciable outflow
wing contribution, although there is a shift in the
line velocity along the direction of the outflow. The CH$_3$OH emission
is centrally concentrated on the IRAS position,  and in the central $200''$
(0.1 pc) it decreases radially in a manner intermediate between 
what would be expected for optically thin emission with density power laws
of $r^{-1}$ and $r^{-1.5}$. The estimated mass in this region is 
5-10 M$_\odot$. No evidence for CH$_3$OH or SiO abundance enhancement is 
found in the outflow.

4. The H$_2$CO(2$_{12}$--1$_{11}$) emission is self absorbed at ambient 
velocities, and presents spectra with brighter blue peak, characteristic of 
inward motions, toward the central $40''$. With a simple model,
we estimate an average infall speed of 0.02 km s$^{-1}$ and an
infall rate of $2\times 10^{-6}$~M$_\odot$ yr$^{-1}$.
At high velocities, the H$_2$CO(2$_{12}$--1$_{11}$)
line presents bright wings in the same sense as
the CO outflow wings, indicative of outflow acceleration. Comparing
the H$_2$CO and CO wing intensities we find that
the H$_2$CO abundance in the outflow regime is enhanced 
with respect to the ambient regime by a factor of 20. 

5. The combination of CO, CH$_3$OH, and H$_2$CO data shows that the L483
outflow is less extreme than other outflows from
Class 0 objects, like L1448-C and L1157, although it has some of their
characteristics, such as gas heating and some abundance 
enhancement. We therefore suggest that the Class 0 source at the
center of the L483 outflow is more evolved than other Class 0 sources,
and it is in its transition to become a Class I object.

\begin{acknowledgements}
This research has made use of the Simbad data base, operated at CDS,
Strasbourg, France, and NASA's Astrophysics Data System Abstract Service.
The Digitized Sky Survey was produced at the Space Telescope Science 
Institute under US Government grant NAG W-2166. 
MT and RB acknowledge partial support from the Spanish
DGESIC grant PB96-104, PCM acknowledges support from NASA Origins of Solar 
Systems grant NAG5-6266, and DM acknowledges partial support from
grant FONDECYT 1990632.
\end{acknowledgements}

\appendix 
\section{Separation between outflow and ambient cloud emission}

Estimating the outflow energetics for L483 is complicated 
because its low velocity and the presence of an extended component. 
In this appendix we discuss the method we have applied to correct for
contamination by {\em background} gas seen in {\em emission}. Unfortunately,
the part of the outflow emission {\em absorbed} by {\em foreground} gas
cannot be recovered by any simple means.

\begin{figure*}
  \resizebox{\hsize}{!}{\includegraphics{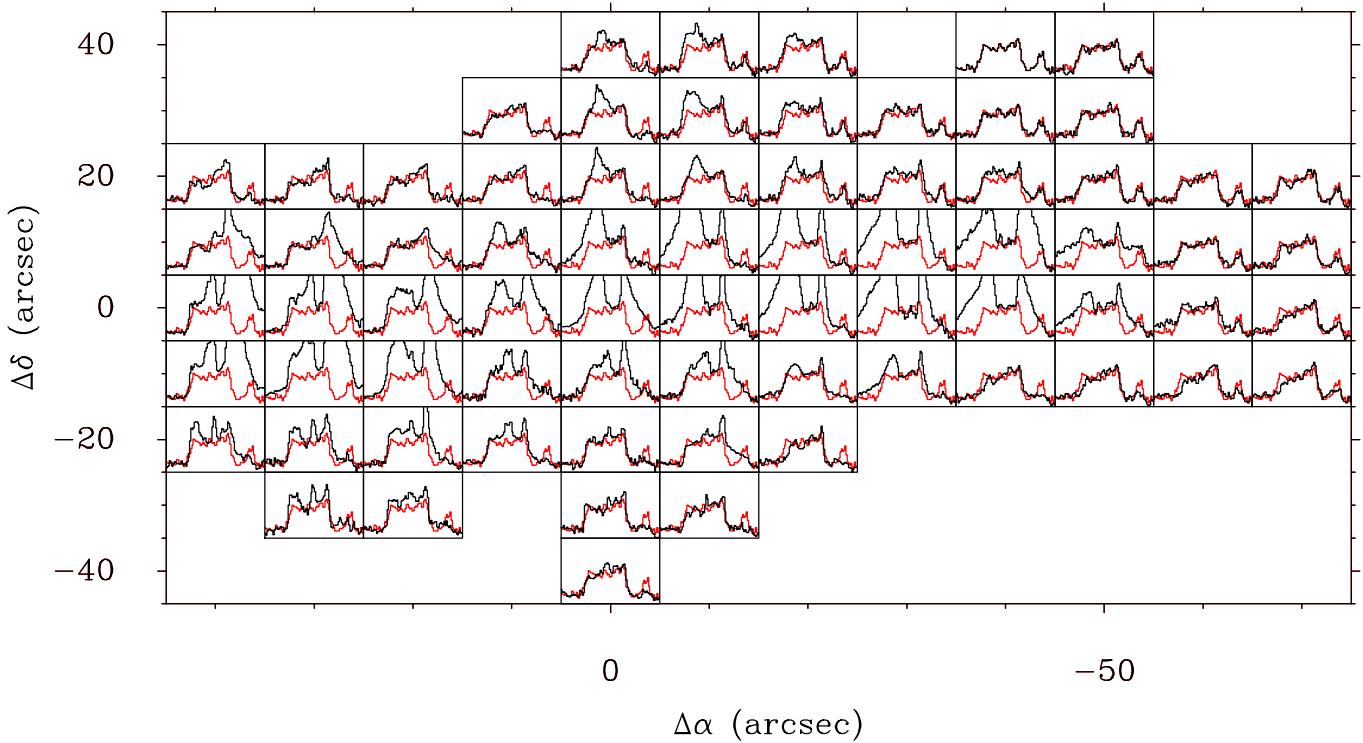}}
  \caption{Map of $^{12}$CO(2--1) spectra towards L483 outflow (dark shade) 
with
the spectrum from ($-40''$, $40''$) superposed for comparison (light shade).
Note how positions with no outflow emission show very similar
spectra. This allows subtracting the emission from gas unrelated to outflow.
For each spectrum, the velocity scale goes from $V_\mathrm{LSR}$ 0 to 10 
km s$^{-1}$
and the $T_\mathrm{mb}$ scale goes from -1.5 to 10 K.
        }
\label{app_fig1}
\end{figure*}

The reason it is possible to correct for the extended emission in L483 
is because this emission seems constant over the flow. This can be seen
in Figure~\ref{app_fig1}, where we present a map of spectra 
with the spectrum from ($-40''$, $40''$) (a representative ambient position) 
superposed
in lighter shade over each map position. As the figure shows, positions without 
outflow wings have spectra with the same shape, suggesting 
that the ambient cloud superposed to the 
outflow emission contributes everywhere with a similar spectrum. If the outflow
emission is optically thin (as suggested by the lack of $^{13}$CO(1--0)
wings), this extended component can be subtracted out, leaving outflow-only
emission as a residual. The origin of the extended component is in part ambient
cloud background to the outflow (the foreground part causes the absorption 
and cannot be corrected for) and in part background emission from 
unrelated gas from the Aquila Rift, as discussed in section 3.1.1. 
The fact that its spectrum is rather 
flat topped suggests that this emission is partly thick with a kinetic
temperature around 10~K, like that of the ambient L483 emission (but
less extreme given its weaker $^{13}$CO(1--0) emission). Given these
characteristics, the gas that does not
appear in absorption is most likely background to the outflow and
therefore susceptible to correction.

\begin{figure}
  \resizebox{\hsize}{!}{\includegraphics{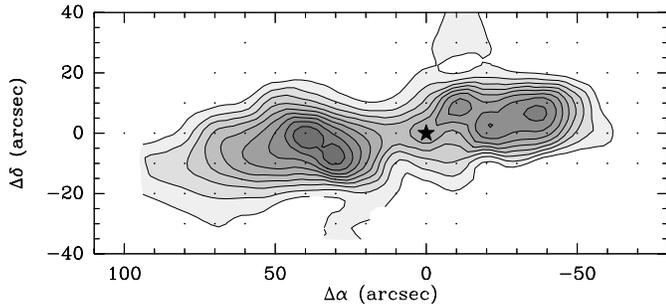}}
  \caption{Integrated $^{12}$CO(2--1) intensity map after background 
emission subtraction (from -2 to 12 km s$^{-1}$). Note how all residual 
emission appears to be from outflow (compare with Figure~\ref{fig2}). 
Lower contour and step are 6 K km s$^{-1}$ and 
offsets are as in Fig.~\ref{fig4}a.
        }
\label{app_fig2}
\end{figure}

To avoid adding noise in the process, we have used as background emission 
the average of all non outflow positions, and we have 
subtracted this spectrum to each observed position. The result seems to
contain outflow emission only, as illustrated by the total integrated 
emission map of Figure~\ref{app_fig2}, which is very similar 
to the outflow map in Fig.~\ref{fig2} (where no background subtraction 
was applied). To these background-subtracted spectra we have applied
the standard energetics analysis (cf. Margulis \& Lada \cite{Mar85}),
and have ignored any contribution in the range $V_\mathrm{LSR}=4$-7~km s$^{-1}$,
as these velocities are contaminated by self absorption. For being forced
to ignore these very low velocities, our estimates are necessarily lower limits
to the real outflow parameters.

\end{document}